\documentclass[prl,letterpaper,twocolumn,showpacs,superscriptaddress]{revtex4}
\usepackage{graphicx,psfrag,amsmath,amssymb,amsfonts,bbm,latexsym,color,dcolumn}

\begin{document}

\title{Quantum limit in resonant vacuum tunneling transducers}

\author{Roberto Onofrio}

\affiliation{Dipartimento di Fisica ``G. Galilei'', Universit\`a di
  Padova, Via Marzolo 8, 35131 Padua, Italy}

\author{Carlo Presilla}

\affiliation{Dipartimento di Fisica, Universit\`a di Roma ``La
  Sapienza'', Piazzale A. Moro 2, 00185 Rome, Italy}

\date{10 May 1993, published in {\sl Europhys. Lett.} {\bf 22} (1993) 333}

\begin{abstract}
We propose an electromechanical transducer based on a
resonant-tunneling configuration that, with respect to the standard
tunnelling transducers, allows larger tunnelling currents while using
the same bias voltage. The increased current leads to a decrease of
the shot noise and an increase of the momentum noise which determine
the quantum limit in the system under monitoring. Experiments with
micromachined test masses at 4.2 K could show dominance of the
momentum noise over the Brownian noise, allowing observation of
quantum-mechanical noise at the mesoscopic scale.
\end{abstract}

\maketitle

Recently a novel electromechanical transducer based upon vacuum
tunnelling of electrons has been proposed to detect displacements of a
macroscopic mass \cite{EPL1}. A variation of the distance between the
test mass and a tip changes the tunnelling current and whenever small
fractions of the current are appreciable, corresponding displacements
of the test mass, which are small fractions of the De Broglie
wavelength of the tunnelling electrons, are also detectable. 
The relevance of this new class of transducers has been emphasized
especially concerning detection of gravitational waves using bar
antennae \cite{EPL1,EPL2}, design of quantum standard of current in
metrology \cite{EPL3} and study of quantum-mechanical noise at the
mesoscopic scale \cite{EPL4}. 

Vacuum tunnelling transducers are intrinsically quantum limited
\cite{EPL5}. The small output capacitance allows to neglect the
back-action noise due to the amplifier following the transducer in the
detection chain with respect to the quantum uncertainties coming from
the tunnelling process in itself. In this last process two
uncorrelated sources of noise have been identified. Firstly, the shot
noise due to the discrete nature of the electric charge is responsible
for a position uncertainty of the test mass. Secondly, the
fluctuations in the momentum imparted by the electrons to the test
mass give rise to a momentum uncertainty of the test mass. The product
of these two quantities is of the order of $\hbar/2$ reaching exactly
this value in the case of a transducer schematized by a square-well
barrier \cite{EPL6}. 

Brownian noise arising from the coupling of the test mass to the
environment usually dominates over the quantum noise and destroys the
quantum properties of the test mass. Suppression of the Brownian noise
contribution is crucial for improving the sensitivity of position
transducers until the standard quantum limit is reached and eventually
surpassed as required in high-precision experiments in general
relativity \cite{EPL7}. Moreover, repeated monitoring at a quantum
level of sensitivity of a single degree of freedom of a macroscopic
mass is relevant to understand quantum measurement theory
\cite{EPL8}. It is therefore important to study mechanisms for which
the quantum noise can be made dominant with respect to the Brownian
noise. In this letter we propose the use of resonant vacuum tunnelling
transducers to achieve such a goal. We will apply the uncertainty
principle to a double barrier in which resonant tunnelling occurs and
we will compare the noise figures to the corresponding non-resonant
case.

Let is consider a tunnelling transducer driven by an incident current
$I$, {\it i.e.} $I$ is the current which should flow in the device if
the tip and the test mass were in contact. Due to this current, during
a sampling time $\Delta t$ the number of electrons which attempt to
tunnel across the vacuum gap, the number of incident electrons
hereafter, is given by
\begin{equation}
N=\frac{I}{e} \Delta t.
\end{equation}
In a first stage we suppose that all the incident electrons have the
same energy $E$, after we will discuss the case of a biased device
with electrons having Fermi distribution. Let $T(E,l)$ be the
transmission coefficient at energy $E$ for a distance $l$ between the
tip and the test mass. A fraction $T$ of the $N$ incident electrons
gives rise to a measured tunnelling current $I_T=TI$. Due to the
discrete nature of the charge carriers a shot noise in the measured
tunnelling current $I_T$ inversely proportional to $\sqrt{N}$ arises. 
The test mass position is inferred by means of the tunnelling current
through the dependence of the transmission coefficient on the distance
$l$ and a shot noise position uncertainty $\Delta \lambda$ for the
test mass also arises \cite{EPL6}:
\begin{equation}
\Delta \lambda^2=\frac{\Delta l^2}{N}=\frac{1}{N}T(1-T)|\frac{\partial
  T}{\partial l}|^{-2}.
\end{equation} 
This uncertainty has been expressed in terms of the uncertainty
$\Delta l$ due to a single electron incident at energy $E$. At the
same time the $N$ incident electrons impart a momentum uncertainty
$\Delta \pi$ to the test mass \cite{EPL6}:
\begin{equation}
\Delta \pi^2=N \Delta p^2,
\end{equation}
where $\Delta p$ is the test mass momentum uncertainty due to a single
electron incident at energy $E$. On the basis of ref. \cite{EPL4}
$\Delta p^2=(J_p^t/J_{in})^2-J_{p^2}^t/J_{in}$, where $J_{in}$ is the
incident electron flux and $J_p^t$ and $J_{p^2}^t$ are the momentum
and the momentum-squared fluxes evaluated in the vacuum zone.

The two quantum noise sources increase the energy of the test mass. If
we schematize the test mass by a harmonic oscillator at rest with mass
$M$ and angular frequency $\omega$ the energy increase in the sampling
time $\Delta t$  will be
\begin{equation}
\Delta \epsilon=\frac{\Delta \pi^2}{2M}+\frac{1}{2}M \omega^2 \Delta \lambda^2.
\end{equation}
This can be considered as the exchange of energy between the test mass
(measured object) and the electrons (meter) due to the quantum
measurement process in the time $\Delta t$. 

Superimposed to the two quantum noise sources there is the Brownian
motion of the test mass coupled to the external environment. Taking
into account also this contribution \cite{EPL9} the total variation of
the test mass energy in the time $\Delta t$ will be
\begin{equation}
\Delta \epsilon=\frac{\Delta \pi^2}{2M}+\frac{1}{2} M \omega^2 \Delta
\lambda^2+\frac{k_B \theta}{Q} \omega \Delta t,
\end{equation}
where $\theta$ is the thermodynamical temperature of the reservoir
schematizing the external environment and $Q$ the mechanical quality
factor expressing the relaxation of the harmonic oscillator in the
thermal bath. The total energy introduced by the quantum measurement
process and by the thermodynamical reservoir may be converted into a
total effective displacement $\Delta \xi$, representing the
sensitivity of the transducer for a measurement of duration $\Delta
t$, through the relation $\Delta \epsilon=M \omega^2 \Delta \xi^2$
\begin{equation}
\Delta \xi^2=\frac{\Delta l^2}{2N}+N\left(\frac{\Delta p^2}{2M^2\omega^2}
+\frac{k_B \theta e}{M \omega Q I}\right).
\end{equation}
In this formula the dependence upon the number of incident electrons
has been emphasized to show that both for small values of $N$, when
the shot noise is dominant, and for large values of $N$, when the sum
of the momentum uncertainty and of the thermal noise is dominant,
large values of the effective displacement are obtained. A minimum
value for the effective displacement will be achieved in an
intermediate situation:
\begin{equation}
\Delta \xi_{opt}^2=\frac{\Delta l \Delta
  p}{M\omega}\sqrt{1+\frac{2k_BM\omega e}{Q\Delta p^2 I}}.
\end{equation}
This optimal sensitivity corresponds to the quantum limit when the
thermal contribution is negligible, {\it i.e.} when 
\begin{equation}
\frac{2 k_B \theta M \omega e}{Q \Delta p^2} I \ll 1.
\end{equation}
For instance in a square-well tunnelling transducer we have exactly
$\Delta l \Delta p = \hbar/2$ \cite{EPL4,EPL6} and when the inequality
(8) holds we get the standard quantum limit $\Delta
\xi_{opt}^2=\hbar/2 M \omega$, for an optimal number of incident
electrons $N_{opt}=\hbar M \omega/2 \Delta p^2$. 

The requirement of thermal noise negligible with respect to quantum
momentum noise is difficult to satisfy for a single-barrier transducer
\cite{EPL4}. Due to the small value of the single-electron momentum
uncertainty $\Delta p$ one has to use a large incident current $I$, a
high mechanical quality factor $Q$, a small test mass $M$ and a low
temperature $\theta$ in order to satisfy the inequality (8). A
different situation arises in a resonant tunnelling transducer. In
fig. 1 we show a possible scheme for a transducer based upon a
double-barrier potential. A double junction or an impurity zone is
grown on the surface of the test mass producing an effective potential
barrier against the current flow. when the tip is put near the test
mass a resonant double barrier is achieved. At a resonance energy the
transmission coefficient $T=4 T_1 T_2/(T_1+T_2)^2$ of the double
barrier may be expressed in terms of the transmission coefficients
$T_1$ and $T_2$ of each single barrier \cite{EPL10}. A variation of
the distance $l$ between the tip and the test mass changes the
transmission coefficient $T_1$. For $T_2 \gg T_1$, as in the
situations we will analyze in the following, the simple relationship
holds
\begin{equation}
\frac{\partial T}{\partial l} \simeq \frac{T}{T_1} \frac{\partial
  T_1}{\partial l}.
\end{equation}
With respect to the single-barrier case we have at resonance a
single-electron position uncertainty smaller by a factor $T/T_1$ and,
because of $\Delta l \Delta p \simeq \hbar/2$, a larger
single-electron momentum uncertainty.

A quantitative comparison of a resonant tunnelling transducer to a
non-resonant one is shown in fig. 2. In the former case around the
resonance energy the position and momentum uncertainties transferred
by a single incident electron to the test mass have a value
respectively smaller and larger than in the corresponding non-resonant
case. At the same time their product remains close to $\hbar/2$. It
should be observed that for both resonant and non-resonant cases the
shape of the momentum uncertainty $\Delta p$ closely follows the shape
of the transmission coefficient $T$. Indeed we can define a momentum
uncertainty per single tunneling electron as $\Delta p_T^2=\Delta
p^2/T \simeq 2 m V_0$ which only depends upon the vacuum barrier $V_0$
and the effective electron mass m. This consideration allows to
rewrite the inequality (8) in terms of the tunnelling current $I_T$
and of $\Delta p_T^2$ instead of the corresponding quantities for the
incident electron
\begin{equation}
\frac{2 k_B \theta M \omega e}{Q \Delta p_T^2 I_T} \ll 1.
\end{equation}
It is evident that this equality is better satisfied for a resonant
configuration because in this case a larger tunnelling current $I_T$
for a given incident current $I$ may be obtained. 

The influence of a Brownian noise source is shown in fig. 3. A
non-negligible Brownian nose with a ratio $\theta/Q=10^{-5}$ K causes
a worsening of the optimal sensitivity $\Delta \xi_{opt}^2$ by a
factor $10^2$ with respect to the case $\theta/Q=0$ in a non-resonant transducer.
On the other hand, in the same conditions a resonant transducer
working at the resonance energy retains an optimal sensitivity close
to the quantum limit.

\begin{figure}[t]
\includegraphics[width=0.95 \columnwidth]{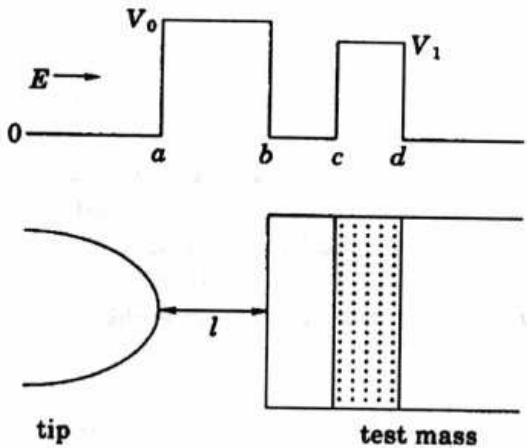}
\caption{Schematic view of a resonant tunnelling transducer and
  corresponding effective one-dimensional double-barrier potential.}
\label{fig1}
\end{figure}

\begin{figure}[t]
\includegraphics[width=0.95 \columnwidth]{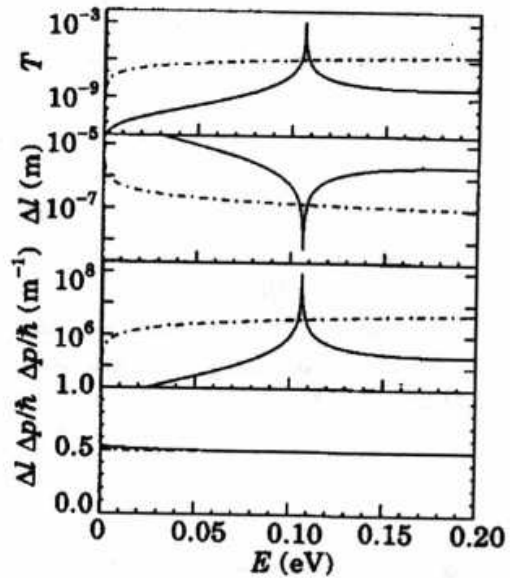}
\caption{Comparison of various quantities {\it vs.} the energy $E$ of the
incident electrons for a resonant double-barrier (solid) and a
single-barrier (dot-dashed) transducer. From the top to the bottom,
respectively: transmission coefficient $T$, position uncertainty
$\Delta l$, momentum uncertainty $\Delta p$, both due to tunnelling of
a single electron, and position-momentum uncertainty product in units
of $\hbar$. The example corresponds to the choice of GaAs tip and test
mass having an AlAs barrier and parameters (see fig. 1) $b-a= 20 \AA$,
$c-b= 50 \AA$, $d-c= 20 \AA$, $V_0=4 eV$, $V_1=1 eV$ and effective
electron mass $m=0.1 m_e$. The single-barrier case is obtained 
with the same parameters except $d-c=0$.}
\label{fig2}
\end{figure}

In a more realistic approach one has to consider not only the effect
of the Brownian noise but also the partial loss of quantum-mechanical
coherence due to elastic scattering. If the escape time of the
electrons from the well $\tau_{esc}$ is longer than the
inelastic-scattering time $\tau_i$ the coherent tunnelling current
decreases and it is only partially compensated by a sequential
tunnelling current \cite{EPL11}. In the situation described in fig. 3
we have $\tau_{esc} \simeq [2(c-b)/v]2/(T_1+T_2)=10^{-11}$ s, where
$v$ is the velocity of the electrons at the resonance. If inelastic
phonon scattering is assumed to dominate, we estimate $\tau_i \simeq
10^{-13}$ s. The loss of coherence can be included by introducing a
phenomenological factor $\gamma$ which is related to the damping of
the wave function due to inelastic scattering and expressed in terms
of $\tau_i$ through the relationship $2 \tau_i=[2(c-b)/v]/(1-\gamma)$
\cite{EPL12}. By using the estimated value of $\tau_i$ in our
configuration, we get $\gamma=0.95$ and the effect of the decoherence
is shown in fig. 3 as a less pronounced peak which still allows for an
order of magnitude improvement when compared to the single-barrier
configuration. Such improvements, both in the fully coherent
configuration ({\it i.e.} $\gamma=1$) and in the presence of inelastic
scattering, shown in fig. 3 are limited to an energy range of the
order of the resonance energy width. 

\begin{figure}[t]
\includegraphics[width=0.95 \columnwidth]{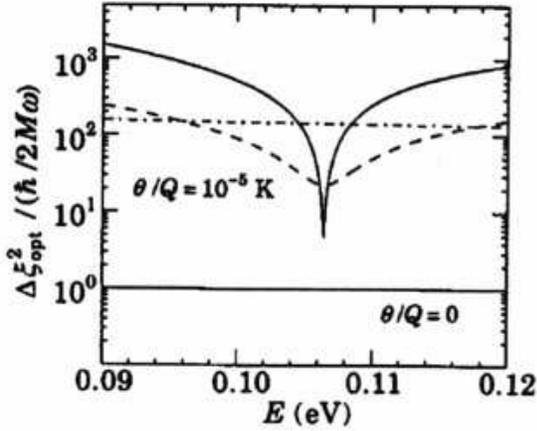}
\caption{
Optimal sensitivity $\Delta \xi_{opt}^2$ {\it vs.} the energy $E$ of
the incident electrons for a resonant double barrier with coherent
tunnelling (solid) and in the presence of sequential tunnelling
(dashed, $\gamma$=0.95) and single-barrier (dot-dashed) transducer in
the case of two different values of the ratio $\theta/Q$ (the two
curves are indistinguishable for $\theta/Q=0$). We have chosen
$M=10^{-10}$ Kg, $\omega=2 \pi 10^5$ s${}^{-1}$, $I=$1 A ($I_T \simeq
10^{-6}$ A and $I_T \simeq 10^{-3}$ A for the single and double
barriers, respectively) and the parameters of the barriers as in fig. 2.}
\label{fig3}
\end{figure}

\begin{figure}[t]
\includegraphics[width=0.95 \columnwidth]{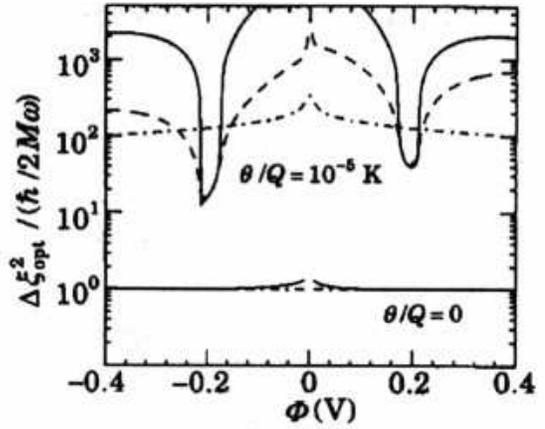}
\caption{
Optimal sensitivity $\Delta \xi_{opt}^2$ {\it vs.} the bias voltage 
$\Phi$ between the tip and the test mass for a resonant double barrier
with coherent tunnelling (solid) and in the presence of sequential tunnelling
(dashed, $\gamma$=0.95) and single-barrier (dot-dashed) transducer in
the case of two different values of the ratio $\theta/Q$.  
We have chosen a Fermi energy $E_F=0.02$ eV, a transverse surface
$S=10^{-9}$ m${}^{2}$ ($I_T \simeq 10^{-6}$ A and $I_T \simeq 10^{-4}$ A
for the single and double barriers, respectively) and all the other 
parameters as in fig. 3.}
\label{fig4}
\end{figure}

Production of ballistic electrons with energy spread smaller than the
resonance energy width is within the current semiconductor technology
capabilities \cite{EPL13}. A similar gain may also be obtained using a
biased device in which a Fermi distribution of electrons gives rise to
the tunneling current. By integrating over the transverse states the
energy distribution of the incident electrons, representing the
differential form of eq. (1), is expressed by \cite{EPL14}
\begin{equation}
\frac{dN}{dE}=\frac{mSk_B\theta}{2 \pi^2 \hbar^3} \ln 
\left[\frac{1+\exp[(E_F-E)/k_B\theta]}
{1+\exp[(E_F-E-e\Phi)/k_B\theta]}\right] \Delta t,
\end{equation}
where $S$ is the transverse surface, $\Phi$ is the bias voltage and
$E_F$ the Fermi energy. This allows to evaluate the integrated shot
noise position uncertainty
\begin{equation}
\Delta \lambda^2=\int_0^\infty \frac{dN}{dE}T(1-T) dE 
{\left[\int_0^\infty \frac{dN}{dE} 
|\frac{\partial T}{\partial l}| dE \right]}^{-2}
\end{equation}
and the integrated momentum uncertainty
\begin{equation}
\Delta \pi^2= \int_0^{\infty} \frac{dN}{dE} \Delta p^2 dE.
\end{equation}
By repeating the same arguments of the monoenergetic case an optimal
sensitivity may be obtained at a proper sampling time:
\begin{equation}
\Delta \xi_{opt}^2=\frac{\Delta \lambda \Delta \pi}{M \omega}
\sqrt{1+\frac{2k_B \theta M \omega}{Q \Delta \pi^2/\Delta t}}.
\end{equation}
The product $\Delta \lambda \Delta \pi$ remains close to $\hbar/2$ and
the quantum limit is reached when the second term inside the square
root is negligible. Due to the constancy of the quantity $\Delta
p_T^2=\Delta p^2/T \simeq 2 mV_0$ this condition is expressed again by
the inequality (10) in terms of the tunnelling current $I_T$,
\begin{equation}
I_T=\frac{e m S k_B \theta}{2 \pi^2 \hbar^3}
\int_0^\infty \ln \left[
\frac{1+\exp[(E_F-E)/k_B \theta]}{1+\exp[(E_F-E-e\Phi)/k_B\theta]} \right]TdE.
\end{equation}
In fig. 4 we show the dependence of the optimal sensitivity as a
function of the bias voltage for the same thermal contributions of
fig. 3 with and without the effect of inelastic scattering. Despite
the integration over all the available electrons the improvement in
the use of the resonant configuration at the proper bias voltage
remains one order of magnitude higher with respect to the
single-barrier situation. Moreover, the optimal sensitivity has a
slight dependence upon the inelastic-scattering processes.

Resonant tunnelling may be relevant for improving the sensitivity of
the tunnelling transducers proposed to detect gravitational waves. In
this class of transducers the sensitivity is limited, apart from $1/f$
noise which strongly depends upon the material used for the test mass
and the tip, by the shot noise. Due to the increased tunnelling
current in a resonant configuration the shot noise is
decreased\footnote{Recent measurements indicate that the shot noise
  in a double barrier is further reduced below the theoretically
  expected value, see \cite{EPL15}}. For the same reason the
momentum noise contribution is enhanced and studies of macroscopic
quantum noise due to the interaction between the electrons and the
test mass are more easily performed. from fig. 3 and 4 it turns out
that experiments for detecting quantum noise at $\theta=4.2$ K seem
also feasible provided that a quality factor $Q \geq 10^6$ at that
temperature can be achieved. Thus experimental studies of the
quantization of a macroscopic degree of freedom of a micromachined
test mass may open interesting prospects in mesoscopic mechanics. In
particular, coherence properties of such single macroscopic
oscillators, {\it e.g.} the creation of distinguishable states of a
harmonic oscillator already proposed in a quantum optics framework
\cite{EPL16}, and its destruction through the influence of the
reservoir can be investigated. 

\noindent
After the submission we have been aware of an experiment performed
with an optical resonator based on tunnelling of frustrated light
which is the optical counterpart of the device we proposed here
\cite{EPL17}.

\acknowledgments
We are grateful to M. F. Bocko for stimulating comments and to
M. B\"uttiker for pointing us out ref. [15]. This work has been
supported by INFN, Italy.

\end{document}